\documentclass[11pt,superscriptaddress,aps,prd]{revtex4}

\everymath{\displaystyle}
\usepackage{graphicx,amsmath,amssymb,amsfonts,amstext,latexsym,stmaryrd,amscd,txfonts,pxfonts,hyperref}
\hypersetup{
    colorlinks=true,
    linkcolor=blue,
    filecolor=magenta,      
    urlcolor=black,
		citecolor=red,
		}

\begin{document}

\title{The irreducible mass of a regular rotating black hole}

\author{F. L. Carneiro}
\email{fernandolessa45@gmail.com}
\affiliation{Universidade Federal do Norte do Tocantins, 77824-838, Aragua\'ina, TO, Brazil}

\author{S. C. Ulhoa}
\email{sc.ulhoa@gmail.com}
\affiliation{Instituto de F\'isica, Universidade de Bras\'ilia, 70910-900, Bras\'ilia, DF, Brazil}
\affiliation{International Center of Physics, Instituto de F\'{\i}sica, Universidade de
Bras\'{\i}lia, 70.910-900, Brasilia, DF, Brazil}
\affiliation{Canadian Quantum Research Center,\\ 
204-3002 32 Ave Vernon, BC V1T 2L7  Canada}

\author{J. F. da Rocha-Neto}
\email{rocha@fis.unb.br}
\affiliation{Instituto de F\'isica, Universidade de Bras\'ilia, 70910-900, Bras\'ilia, DF, Brazil}

\author{J. W. Maluf}
\email{jwmaluf@gmail.com}
\affiliation{Instituto de F\'isica, Universidade de Bras\'ilia, 70910-900, Bras\'ilia, DF, Brazil}

%%%%%%%%%%%%%%%%%%%%%%%%%%%%%%%%%%%%%%%%%%%%%%%%%%%%%%%%%%%%%%%%%%%%%%%%%%%%%%%%%%%%%
\begin{abstract}
This article presents an analysis of regular rotating black hole solutions within the framework of Teleparallel Equivalent to General Relativity (TEGR). The study evaluates the total energy and derives an analytical expression for the irreducible mass of a regular black hole. The results reveal the significance of these regular black holes as approximations of real astrophysical objects. The investigation explores the behavior of the total energy for different surfaces and its value at spatial infinity. Additionally, the article addresses the instability of the inner horizon and examines the inertial acceleration of an observer inside the inner horizon.

\end{abstract}
%%%%%%%%%%%%%%%%%%%%%%%%%%%%%%%%%%%%%%%%%%%%%%%%%%%%%%%%%%%%%%%%%%%%%%%%%%%%%%%%%%%%%

\maketitle

\date{\today}

%%%%%%%%%%%%%%%%%%%%%%%%%%%%%%%%%%%%%%%%%%%%%%%%%%%%%%%%%%%%%%%%%%%%%%%%%%%%%%%%%%
\section{Introduction}
%%%%%%%%%%%%%%%%%%%%%%%%%%%%%%%%%%%%%%%%%%%%%%%%%%%%%%%%%%%%%%%%%%%%%%%%%%%%%%%%%%%

The forty-eight-year-old search for exact solutions of Einstein's gravitational theory culminated with the discovery by Roy Kerr in 1963 \cite{kerr1963gravitational} of the now-known Kerr solution. This solution represents the spacetime geometry of a rotating black hole, where the parameters $m$ and $a$ are related to its mass and angular momentum per unit mass, respectively. It belongs to the Petrov type D class. The Kerr solution encompasses two coordinate singularities, corresponding to two null horizons: an event horizon $r_{+}$ and a Cauchy horizon $r_{-}$ \cite{torres2022regular}. These singularities can be eliminated through an appropriate coordinate transformation.

The Kerr solution is more general than the Schwarzschild solution and provides a more realistic description of physical systems, considering that celestial bodies are observed to rotate. With the aid of the Newman-Janis algorithm \cite{newman1965note}, the Kerr solution quickly became one of the most widely used solutions in general relativity. It has been included in textbooks as an example of a standard solution of the theory. Although more realistic than the solution for a static black hole, the Kerr solution still contains a curvature singularity in the form of a ring singularity located in the equatorial plane $\theta=\pi/2$. Unlike the coordinate singularities that can be removed by a change of coordinates, this singularity is of a geometric nature allowed by general relativity. It is important to note that this singularity is a scalar singularity, meaning that it cannot be eliminated through a change of coordinates.

The worldline of a particle does not necessarily need to terminate at the singularity, as is the case for the Schwarzschild solution. However, the existence of physical curvature singularities has profound consequences for our understanding of general relativity (GR) and the nature of gravity itself \cite{joshi2014spacetime}. These consequences range from the cosmological censorship proposed by Penrose \cite{penrose1969gravitational} in both its weak and strong forms to the incompleteness of general relativity, suggesting that GR may break down in extreme scenarios. Consequently, the existence of singularities is viewed by some as a problem within the theory and a point of incompatibility with quantum mechanics \cite{marolf2017black}, as the unique information of a particle is inevitably lost upon reaching the singularity.

However, it is important to note that the mathematical possibility of a feature in a theory does not necessarily imply its existence in nature, as physical theories provide approximate descriptions of natural phenomena. For example, it is possible to modify Maxwell's electrodynamics to allow for magnetic monopoles, or even conceive of negative mass within Newton's theory. Therefore, some authors advocate for modifications not to the theory itself, but to the solutions, aiming to find solutions that represent physical phenomena without curvature singularities \cite{ayon2005four,ayon2000bardeen,balart2014regular}. In these solutions, it is common to introduce non-traditional sources to avoid the presence of a singularity. On the experimental side, advancements in gravitational wave astronomy have drawn attention to the possibility of detecting real black hole mimickers \cite{mazza2021novel}, thereby increasing the relevance of investigating such geometries.

The regularization of black holes began with the Schwarzschild solution. In 1968, Bardeen considered a charged matter collapse that resulted in a non-singular configuration with a matter core replacing the singularity \cite{bardeen1968non}. In the core of the black hole, the source exhibits repulsive behavior, preventing the occurrence of a singular collapse. The source remains covered by an event horizon, maintaining the exterior geometry similar to that of the Schwarzschild solution. However, the possible energy-momentum tensor of the source violates the weak energy condition \cite{rodrigues2018bardeen, zaslavskii2010regular}. Subsequently, other authors proposed various types of regular black holes, including the Hayward one, featuring a dynamic event horizon with a high-density core \cite{hayward2006formation}. In all of these solutions, regularization is achieved through the specification of a mass function $m(r)$.

When applying the Newman-Janis algorithm to the regular black holes of Bardeen and Hayward, a regular rotating black hole solution is obtained. Each distinct solution can be obtained by specifying the mass function, all without singularities. However, these solutions violate the weak energy condition when the rotation parameter $a=0$. To address this issue, Bambi and Modesto proposed a model with minimal violation of the weak energy condition \cite{bambi2013rotating}. Nevertheless, any deviation from spherical symmetry leads to the violation of the weak energy condition for "any reasonable choice of the mass function" \cite{neves2014regular}.

While energy conditions have some significance, they are not fundamental physical laws of nature. Their absence does not necessarily invalidate a solution. However, constructing solutions by changing the mass function may introduce an inner horizon with non-zero surface gravity. Perturbations near this horizon can lead to an exponential growth of energy, known as the mass inflation instability \cite{poisson1989inner,ori1991inner,hamilton2010physics}. Carballo-Rubio and collaborators addressed this issue by constructing a geometry without the mass inflation problem \cite{carballo2022regular}. Following their steps, Franzin and collaborators constructed a geometry using a conformal factor to regularize the black hole and a mass function to stabilize the inner horizon \cite{franzin2022stable}. In this paper, we will consider their solution and review its aspects in section \ref{review}.

Continuing the exploration of regular rotating black holes, we delve into one of the most intriguing applications of a rotating black hole: the energy extraction through the Penrose process \cite{penrose1969gravitational}. In 1971, Penrose proposed that a particle plunging into the ergosphere of a Kerr black hole could undergo disintegration, resulting in two new particles. One of these particles would plunge into the event horizon with negative energy, while the other would escape the ergosphere with greater energy than the initial particle. This process enables the extraction of rotational energy from a Kerr black hole \cite{penrose1971extraction}. It naturally raises the question of how much energy can be extracted from the black hole by the particle. If a Kerr black hole initially possesses an energy $E$, and the maximum amount of energy extracted is denoted as $\epsilon_{\text{max}}$, then the remaining energy of the black hole after the extraction is given by $E_f$, where $E_f = E - \epsilon_{\text{max}}$. This residual energy is often referred to as the remaining energy $E_f$ after the extraction of the maximum energy $\epsilon_{\text{max}}$ that can be extracted from the black hole. Consequently, the mass associated with this irreducible energy $E_{f}$ is termed the irreducible mass of the Kerr black hole.
Expanding upon this concept and leveraging the previous result for the energy of a particle in geodesic motion with a turning point \cite{carter1968global}, Christodoulou considered an equatorial orbit that satisfies the lowest possible energy for the in-falling particle and obtained the expression for the irreducible energy of the Kerr black hole \cite{christodoulou1970reversible}
\begin{equation}\label{irr_chris}
m_{irr}=\frac{1}{2}\sqrt{r_{+}^{2}+a^{2}}\,.
\end{equation}
The irreducible mass of a black hole corresponds to the energy that cannot be extracted from it. As nothing can escape the event horizon of a Kerr black hole, we can interpret the irreducible mass as the energy contained within the event horizon $r_{+}$.

Christodoulou's result proved to be very interesting and potentially applicable in the future. Although the irreducible mass is a property of the black hole itself, Christodoulou's method modeled it using the motion of an equatorial particle. To evaluate the irreducible mass based on the black hole's own energy, we need a proper definition of an energy-momentum tensor for the gravitational field. The metric formulation of General Relativity does not provide a true energy-momentum tensor, meaning a tensor that cannot be made zero by a specific choice of coordinates. Modern gravitational theories rely on coordinates to describe real fields, but the existence of a field should not depend on coordinates. Therefore, the results obtained from these pseudo-tensors can only have physical meaning in asymptotically flat spacetimes when integrated over the entire space, as in the case of the ADM energy \cite{arnowitt2008republication}. In an attempt to establish an energy-momentum complex that yields a true energy-momentum tensor, M\o{}ller concluded that any metric formulation of General Relativity makes the establishment of a true tensor infeasible and that a tetrad approach is necessary \cite{moller1964conservation}.

One formulation of the gravitational field that yields equivalent field equations to General Relativity (GR) is the Teleparallel Equivalent of General Relativity (TEGR), which is based on a tetrad framework. In this formulation, a well-defined energy-momentum tensor for the gravitational field can be constructed \cite{maluf2013teleparallel}. This energy-momentum tensor allows us to explore the distribution of energy within the spacetime described by the gravitational field.
Within the event horizon, the energy is effectively trapped, confined to the gravitational grip of the black hole. It is this confinement of energy that gives rise to the concept of irreducible mass.
Given that the energy confined within the event horizon of a black hole cannot escape, it is natural to investigate the relationship between this confined energy and the parameters characterizing the black hole. It has been found that for Schwarzschild black holes, the energy calculated at the horizon is directly proportional to the black hole's mass parameter $M$, with a factor of two. This relationship, expressed as $E=2M$ in geometrized units \cite{martin}, provides a meaningful connection between the trapped energy within the event horizon and the intrinsic mass of the black hole.
Hence, the identification $E_{\text{irr}}=2M_{\text{irr}}$ emerges as a fundamental result in the study of black holes, serving as a crucial foundation for understanding the concept of irreducible mass. It highlights the intimate connection between the trapped energy and the fundamental properties of the black hole, shedding light on the intrinsic nature of these cosmic objects.
Following this procedure the irreducible mass of the Kerr black hole was obtained as \cite{maluf1996gravitational}
\begin{equation}\label{irr_maluf}
E_g\;=\;m\biggl[ {\sqrt{2p}\over 4}+\,
\,{{6p-k^2}\over 4k}ln \,\biggl(
{{\sqrt{2p} + k}\over p}\biggr) \biggr]\;,
\end{equation}
where $p=1+\sqrt{1-k^2}$ and $a=k\,m$. Despite equation (\ref{irr_maluf}) appearing to be distinct from equation (\ref{irr_chris}), it was shown in Ref. \cite{maluf1996gravitational} that their numerical values are very similar.

Ergo, given the reasonable assumption of the existence of regular rotating black holes as good approximations of real astronomical objects \cite{narzilloev2023regular,kumar2020testing,berry2020photon}, and using the method presented in Ref. \cite{maluf1996gravitational}, this article aims to evaluate the total energy (gravitational plus source) of the regular rotating black hole introduced in Ref. \cite{franzin2022stable}. We also seek to find an analytical expression for the irreducible mass of the regular Kerr black hole. Our derived expression reproduces the Kerr limit when the $e$ and $b$ parameters are set to zero, i.e., the limiting case (\ref{irr_maluf}) of Ref. \cite{maluf1996gravitational}. Additionally, we investigate the behavior of the total energy for different surfaces and its value at spatial infinity. Addressing the instability of the inner horizon, we also examine the black hole energy at this surface when the stabilization parameter $e$ is disregarded.

This article is organized as follows. In Section \ref{terg}, we provide a review of some important aspects of TEGR necessary for understanding the method. In Section \ref{review}, we present the regular black hole considered in this article and recapitulate some of its properties. In Section \ref{ge}, we analytically evaluate the total energy density and numerically integrate it for several surfaces with different radii. In Section \ref{mass}, we present our main result, which is the determination of the irreducible energy, and consequently the irreducible mass, of a regular rotating black hole by taking the limit of the event horizon. Finally, in Section \ref{conclusions}, we present our conclusions.

Throughout this article, we use the following notation. $SO(3,1)$ indices are represented by Latin letters at the beginning of the alphabet $a$, $b$, $\ldots$, and local coordinates are indicated by $(0)$, $(1)$, $\ldots$. Greek letters $\mu$, $\nu$, $\ldots$ denote spacetime indices, while $0$, $1$, $\ldots$ denote coordinate indices. All indices run from $0$ to $3$, with time and space indicated by $(0)$, $0$, and $(i)$, $i$, respectively. $SO(3,1)$ indices are raised and lowered using the flat Minkowski tensor $\eta_{ab}$, and spacetime indices are raised and lowered using the metric tensor $g_{\mu\nu}$.

%%%%%%%%%%%%%%%%%%%%%%%%%%%%%%%%%%%%%%%%%%%%%%%%%%%%%%%%%%%%%%%%%%%%%%%%%%%%%%%%%%%%%%%%%%%%%%%%%%%%%%%%%%%%%%%%%%%%%%%%%%%%%%%
%%%%%%%%%%%%%%%%%%%%%%%%%%%%%%%%%%%%%%%%%%%%%%%%%%%%%%%%%%%%%%%%%%%%%%%%%%%%%%%%%%%%%%%%%%%%%%%%%%%%%%%%%%%%%%%%%%%%%%%%%%%%%%%
\section{Teleparallel Equivalent to General Relativity (TEGR)}\label{terg}

Modern gravitational theories consider the description of gravitational phenomena as the emergence of a non-trivial geometry on a manifold. Therefore, distinct geometries can lead to the same gravitational effect, depending on the field equations of the theory. To establish a geometry on a manifold, we need to specify three quantities: the metric $g_{\mu\nu}$, torsion $T_{\lambda\mu\nu}$, and non-metricity $Q_{\lambda\mu\nu}$. For example, by choosing zero torsion and metricity, we have Riemannian geometry.

Let us consider a tetrad field $e^{a}\,_{\mu}$, where $a$ is a local Lorentz index (of the tangent space to the event $x^{\mu}$), and $\mu$ is a spacetime index. With the aid of the tetrads $e^{a}\,_{\mu}$, we can project a spacetime quantity into the tangent space, for instance, $v^{a}=e^{a}\,_{\mu}v^{\mu}$. We can construct a geometry that establishes the parallel transport of the tetrads
\begin{equation}\label{eq3}
\nabla_{\mu}e^{a}\,_{\nu}=\partial_{\mu}e^{a}\,_{\nu} - \Gamma^{\lambda}\,_{\mu\nu}\,e^{a}\,_{\lambda}=0\,,
\end{equation}
i.e., choose the manifold guaranteeing (\ref{eq3}) on the spacetime. Isolating the connection in (\ref{eq3}), we obtain the connection that guarantees the absolute parallelism
\begin{equation}\label{eq4}
\Gamma^{\lambda}\,_{\mu\nu}=e_{a}\,^{\lambda}\partial_{\mu}e^{a}\,_{\nu}\,.
\end{equation}
The afine connection (\ref{eq4}) transports a vector between two distant points without changing its direction. Therefore, the vector is parallel transported. This connection yields a zero curvature tensor. Also,
connection (\ref{eq4}) is not symmetric under the permutation of the lower indices, resulting in a non-zero torsion tensor
\begin{equation}\label{eq5}
T^{\lambda}\,_{\mu\nu}=\Gamma^{\lambda}\,_{\mu\nu}-\Gamma^{\lambda}\,_{\nu\mu}\,.
\end{equation}
The connection (\ref{eq4}) is known as the Weitzenböck connection, and the torsion tensor (\ref{eq5}), along with the metricity condition $Q_{\lambda\mu\nu}=0$, defines the geometry on which the Teleparallel Equivalent to General Relativity (TEGR) is built.

The Lagrangian density of TEGR is constructed using invariants derived from the torsion tensor (\ref{eq5}). First, we substitute (\ref{eq4}) into (\ref{eq5}), resulting in
\begin{equation}\label{eq6}
T^{a}\,_{\mu\nu} = \partial_{\mu}e^{a}\,_{\nu} - \partial_{\nu}e^{a}\,_{\mu}\,.
\end{equation}
By defining the superpotential 
\begin{equation}\label{eq7}
\Sigma^{abc}\equiv\frac{1}{4}\left(T^{abc}+T^{bac}-T^{cab}\right)+\frac{1}{2}\left(\eta^{ac}T^{b}-\eta^{ab}T^{c}\right)\,,
\end{equation}
where $T^{c}\equiv T^{\mu}\,_{\mu}\,^{c}$, we construct the Lagrangian density
\begin{equation}\label{eq8}
\mathcal{L}=-ke\Sigma^{abc}T_{abc}-\mathcal{L}_{M}\,,
\end{equation}
where $\mathcal{L}_{M}$ represents the Lagrangian of the matter-radiation fields, and $k=1/16\pi$.  It is worth noting that the Lagrangian density is invariant under coordinate transformations and global Lorentz transformations. The addition of a total divergence may also make it invariant under local Lorentz transformations. Although the field equations derived from it exhibit the same symmetries as Einstein's equations, precisely because they are equivalent, the tensorial quantities defined here are covariant only under the symmetries of the Lagrangian density itself.
From (\ref{eq8}), we derive the field equations of TEGR \cite{maluf2013teleparallel}
\begin{equation}\label{eq9}
e_{a\lambda}e_{b\mu}\partial_{\nu}\left(e\Sigma^{b\lambda\nu}\right)-e\left(\Sigma^{b\nu}\,_{a}T_{b\nu\mu}-\frac{1}{4}e_{a\mu}T_{bcd}\Sigma^{bcd}\right)=\frac{1}{4k}eT_{a\mu}\,,
\end{equation}
where $T_{a\mu}=e_{a}\,^{\lambda}T_{\lambda\mu}$, and $e$ is the determinant of the tetrads (not to be confused with the stabilization parameter of the black hole).
The field equations (\ref{eq9}) can be written as $R_{a\mu}-\frac{1}{2}e_{a\mu}R=kT_{a\mu}$. Hence, since $R_{\mu\nu}=e^{a}\,_{\mu}R_{a\nu}$,the field equations of TEGR are equivalent to those of GR.

Although (\ref{eq9}) are dynamically equivalent to Einstein's equations, considering the tetrads as the fundamental variable of gravity allows us to express the field equations in a more fundamental form than that of GR. Following the procedure of \cite{maluf2013teleparallel}, we can rewrite equation (\ref{eq9}) as
\begin{equation}\label{eq10}
\partial_{\nu}\left(e\,\Sigma^{a\lambda\nu}\right)=\frac{1}{4k}e\,e^{a}\,_{\mu}\left(t^{\lambda\mu}+T^{\lambda\mu}\right)\,,
\end{equation}
where we define the quantity
\begin{equation}\label{eq11}
t^{\lambda\mu}=k\left(4\Sigma^{bc\lambda}T_{bc}\,^{\mu}-g^{\lambda\mu}\Sigma^{bcd}T_{bcd}\right)\,.
\end{equation}
Noticing that $\Sigma^{a\lambda\nu}$ is antisymmetric in $\lambda$ and $\mu$, differentiating (\ref{eq10}) with respect to $\lambda$, we obtain
\begin{equation}\label{eq12}
\partial_{\lambda}\left[ee^{a}\,_{\mu}\left(t^{\lambda\mu}+T^{\lambda\mu}\right)\right]=0\,.
\end{equation}
Integrating equation (\ref{eq12}), we obtain the continuity equation
\begin{equation}\label{eq13}
\frac{d}{dx^{0}}\int_{V}{d^{3}x\left[ee^{a}\,_{\mu}\left(t^{0\mu}+T^{0\mu}\right)\right]}=-\oint_{S}{dS_{i}\left[ee^{a}\,_{\mu}\left(t^{i\mu}+T^{i\mu}\right)\right]}\,.
\end{equation}
Hence, by considering the surface at the right-hand side of (\ref{eq13}) to be at infinity, we have the conservation of the quantity
\begin{equation}\label{eq14}
P^{a}\equiv\int_{V}{d^{3}x\,e\,e^{a}\,_{\mu}\left(t^{0\mu}+T^{0\mu}\right)}\,.
\end{equation}
The above quantity is conserved within a surface when there is no energy-momentum flux along this surface, e.g., at infinity for asymptotically flat solutions. If we choose a surface with a non-zero energy-momentum flux, the time derivative in (\ref{eq13}) will be non-zero (for an example on the FLRW metric in the dynamical horizon, see \cite{rocha}).
Therefore, we can identify (\ref{eq11}) as the energy-momentum tensor of the gravitational field, and the total energy-momentum enclosed by the surface $S$ as
\begin{equation}\label{eq15}
P^{a}=4k\int_{V}{d^{3}x \, \partial_{i}(e\Sigma^{a0i}})=4k\oint_{S}dS_{i}(e\Sigma^{(0)0i})\,,
\end{equation}
where $S$ is the closed surface enclosing $V$. The energy-momentum (\ref{eq15}) transforms as a 4-vector under global Lorentz transformation, is invariant under change of coordinates of the three dimensional space and also invariant under time reparametrizations.

%Also, (\ref{eq15}) does not transform covariantly under local Lorentz transformations, as expected since such quantity is a spatially integrated one, i.e., dependent on the parameters and, at most, at the coordinate time. The terms of local Lorentz transformation are functions of the coordinates. Hence, a covariance under local Lorentz transformations would make the energy definition dependent on the coordinates.

In this section, we have only provided a brief overview of some important aspects of TEGR for the analysis in this article.
Although we obtained the energy-momentum (\ref{eq15}) through the Lagrangian formalism of TEGR, the same results can be formally derived using the Hamiltonian formulation of the theory. For a comprehensive review, we recommend referring to the Ref. \cite{maluf2013teleparallel}.

%%%%%%%%%%%%%%%%%%%%%%%%%%%%%%%%%%%%%%%%%%%%%%%%%%%%%%%%%%%%%%%%%%%%%%%%%%%%%%%%%%%%%%%%%%%%%%%%%%%%%%%%%%%%%%%%%%%%%%%%%%%%%%%
%%%%%%%%%%%%%%%%%%%%%%%%%%%%%%%%%%%%%%%%%%%%%%%%%%%%%%%%%%%%%%%%%%%%%%%%%%%%%%%%%%%%%%%%%%%%%%%%%%%%%%%%%%%%%%%%%%%%%%%%%%%%%%%
\section{A regular rotating black hole}\label{review}

The Kerr solution in Boyer-Lindquist coordinates are described by
\begin{equation}\label{eq16}
ds^2=
-{{\psi^2}\over {\rho^2}}dt^2-{{2\chi\sin^2\theta}\over{\rho^2}}
\,d\phi\,dt
+{{\rho^2}\over {\Delta}}dr^2 
+\rho^2d\theta^2+ {{\Sigma^2\sin^2\theta}\over{\rho^2}}d\phi^2\,,
\end{equation}
where
\begin{align}
\Delta&= r^2+a^2-2mr\,,  \nonumber \\
\rho^2&= r^2+a^2\cos^2\theta \,,  \nonumber \\
\Sigma^2&=(r^2+a^2)^2-\Delta\, a^2\sin^2\theta\,,  \nonumber \\
\psi^2&=\Delta - a^2 \sin^2\theta\,, \nonumber \\
\chi &=2amr\,.\nonumber
\label{23}
\end{align}
The horizons occur when $\Delta=0$. Therefore, we have the outer (event) horizon at
\begin{equation}\label{eq17}
r_{+}=m+\sqrt{m^{2}+a^{2}}
\end{equation}
and the inner (Cauchy) horizon at
\begin{equation}\label{eq18}
r_{-}=m-\sqrt{m^{2}+a^{2}}\,.
\end{equation}
The coordinate singularities arising from $\Delta=0$ can be eliminated by choosing double null coordinates, which represent the ingoing or outgoing geodesics of light rays.
From the curvature tensor, we can construct 14 scalar invariants \cite{weinberg1972gravitation}, such as the Ricci scalar $R=R^{\mu\nu}R_{\mu\nu}$ and the Kretschmann scalar $\mathcal{K}=R^{\mu\nu\alpha\beta}R_{\mu\nu\alpha\beta}$. When a scalar invariant diverges as it is approached by any (incomplete) curve, the spacetime contains a scalar curvature singularity.

For the Kerr solution (\ref{eq16}), the Kretschmann scalar yields \cite{henry2000kretschmann}
\begin{equation}\label{eq19}
\mathcal{K} = \frac{48m^{2}}{ (r^{2}+a^{2} cos^{2}\theta)^{6}}\Big(r^{6}-15a^{2}r^{4}cos^{2}\theta +15 a^{4} r^{2} cos^{4}\theta-a^{6}cos^{6}\theta\Big)\,.
\end{equation}
A singularity in the scalar (\ref{eq19}) occurs at $r=0$ and $\theta=\pi/2$, indicating the presence of a ring singularity in the equatorial plane of the Kerr solution.
To regularize a black hole, a straightforward approach is to consider $m=m(r)$, as done in the Bardeen regularization of the Schwarzschild black hole \cite{bardeen1968non}. However, not every choice of $m(r)$ can successfully regularize the solution. The regularization is achieved only if all second-order invariants of the curvature tensor remain finite at the ring singularity. This condition is satisfied when \cite{torres2017regular}
\begin{equation}
m(0)=m'(0)=m''(0)=0\,,
\end{equation}
where the prime represent differentiation with respect to $r$.
However, the procedure of solely adopting $m=m(r)$ yields a black hole that possesses a non-zero surface gravity at the inner horizon. Consequently, this leads to an accumulation of energy around the inner horizon for any small perturbation, resulting in the inner horizon becoming a surface with infinite curvature. This phenomenon is known as the mass-inflation problem \cite{ori1991inner}.
Therefore, an alternative approach must be employed to obtain a more realistic solution.

Another method of regularizing a black hole involves introducing a conformal factor. In the following, we present the procedure outlined in Ref. \cite{franzin2022stable}. Starting from the Kerr line element (\ref{eq16}), we define another line element $d\bar{s}^{2}$ that only differs by a conformal factor. That is,
\begin{equation}\label{eq21}
d\bar{s}^{2}=\frac{\Psi}{\rho^{2}}ds^2=\frac{\Psi}{\rho^{2}}\Bigg(
-{{\psi^2}\over {\rho^2}}dt^2-{{2\chi\sin^2\theta}\over{\rho^2}}
\,d\phi\,dt
+{{\rho^2}\over {\Delta}}dr^2 
+\rho^2d\theta^2+ {{\Sigma^2\sin^2\theta}\over{\rho^2}}d\phi^2\Bigg)\,.
\end{equation}
The factor
\begin{equation}
\Psi=\rho^{2}+\frac{b}{r^{2z}}\,,
\end{equation}
where $b$ and $z$ are a parameter and a positive constant, respectively. For the line element (\ref{eq21}), the Ricci scalar yields 
\begin{equation}
R=-\frac{6b\,r^{2z}}{r^{2}\rho^{4}(b^{2}+r^{2z}\rho^{2})^{3}}f(r,\theta)\,,
\end{equation}
where $f(r,\theta)$ is a function that approaches zero faster than or equal to $\rho^{4}$ as the inner ring singularity is approached. The factor $\Psi$ can resolve the scalar singularity problem in the Ricci tensor if $b\neq 0$, and it can also address other scalar invariants.
The parameter $b$ is a new parameter required for metric regularization, while the parameter $z$ can be chosen such that the limit $\displaystyle{\lim_{(r,\theta)\rightarrow(0,\pi/2)}}R$ exists. The smallest value of $z$ that yields a well-defined limit is $z=3/2$ \cite{franzin2022stable}. Thus, we will consider the conformal factor as
\begin{equation}\label{eq24}
\Psi=\rho^{2}+\frac{b}{r^{3}}\,,
\end{equation}
where the parameter $b$ assumes the dimension of $[r]^{5}$.

Although the spacetime described by the line element (\ref{eq21}) is regular even for $m=M=\textit{constant}$, its inner horizon exhibits a non-zero surface gravity, leading to the aforementioned mass-inflation problem. To eliminate the non-zero surface gravity, a function $m=m(r)$ can be selected. There are several choices for non-singular functions. In this article, following Ref. \cite{franzin2022stable}, we choose the function
\begin{equation}\label{eq25}
m(r) = M\,\frac{r^2+\alpha r + \beta}{r^2 + \gamma r+ \mu}\,,
\end{equation}
where $M$ is the "mass" parameter of the Kerr black hole, and $\alpha$, $\beta$, $\gamma$, and $\mu$ are constants that can be determined based on the positions of the horizons, such as
\begin{align}
\alpha &=\frac{a^4+r_-^3 r_+ - 3 a^2 r_- (r_- + r_+)}{2 a^2 M}\,,\label{eq26}\\
\beta &=\frac{a^2 (2 M - 3 r_- - r_+) + r_-^2 (r_- + 3 r_+) }{2 M}\,,\label{eq27}\\
\gamma &= 2M - 3r_- - r_+\,,\label{eq28}\\
\mu &=\frac{r_-^3 r_+}{a^2}\,.\label{eq29}
\end{align}
By determining the positions of the horizons and $M$, the constants can be determined accordingly. Therefore, the mass function introduces only one additional parameter to the solution. Choosing the outer horizon to be similar to that of Kerr (\ref{eq17}), we have
\begin{equation}\label{eq30}
r_{+}=M+\sqrt{M^{2}+a^{2}}
\end{equation}
and the inner one as
\begin{equation}\label{eq31}
r_{-}=\frac{a^{2}}{M+(1-e)\sqrt{M^{2}-a^{2}}}\,,
\end{equation}
where
\begin{equation}\label{eq32}
-3 - \frac{3M}{\sqrt{M^2 - a^2}} < e < 2\,.
\end{equation}
By employing the boundaries (\ref{eq32}), we guarantee that the inner horizon is fully covered by the outer horizon, and that $m(r)$ has no poles. With the choice (\ref{eq30}), the exterior geometry of the solution is expected to be similar to that of the Kerr black hole.

In the subsequent results of this paper, we consider the line element (\ref{eq21}) with $\Psi$ given by (\ref{eq24}), and the horizons given by (\ref{eq30}) and (\ref{eq31}). This geometry represents a regular rotating black hole without closed time-like curves and with a stable inner horizon \cite{franzin2022stable}, and we will evaluate its energy contained in $r_{+}$ which can be identified with the irreducible mass. The metric depends on four parameters: the mass parameter $M$, the angular momentum parameter $a$, the regularization parameter $b$, and the Kerr deviation parameter $e$.

%%%%%%%%%%%%%%%%%%%%%%%%%%%%%%%%%%%%%%%%%%%%%%%%%%%%%%%%%%%%%%%%%%%%%%%%%%%%%%%%%%%%%%%%%%%%%%%%%%%%%%%%%%%%%%%%%%%%%%%%%%%%%%%
%%%%%%%%%%%%%%%%%%%%%%%%%%%%%%%%%%%%%%%%%%%%%%%%%%%%%%%%%%%%%%%%%%%%%%%%%%%%%%%%%%%%%%%%%%%%%%%%%%%%%%%%%%%%%%%%%%%%%%%%%%%%%%%
\section{The gravitational energy}\label{ge}

The energy of a configuration of a gravitational field can be evaluated from the zeroth component of equation (\ref{eq15}). To do so, we need to calculate the torsion tensor (\ref{eq5}) using a set of tetrads associated with the metric (\ref{eq21}).

For every spacetime point described by the coordinates $x^{\mu}$, the tetrads that may be associated to the observer's instantaneous rest frame, consisting of orthonormal vectors $\{e_{(0)}\,^{\mu},e_{(1)},^{\mu},e_{(2)}\,^{\mu},e_{(3)}\,^{\mu} \}$. The observer is assumed to be at rest in their own frame, so $e_{(0)}\,^{\mu}$ must be tangent to the observer's world line, and we can identify it with their four-velocity $u^{\mu}$, i.e., $u^{\mu}=e_{(0)}\,^{\mu}$.

The metric tensor $g_{\mu\nu}$ can be constructed from the tetrads $e_{\mu\nu}$ using the relation $g_{\mu\nu}=e^{a}\,_{\mu}e_{a\nu}$. The metric tensor has 10 independent components, while the tetrads have 16 components. The additional six degrees of freedom correspond to the inertial state of the observer. From the torsion tensor (\ref{eq5}), we can construct the antisymmetric acceleration tensor \cite{maluf2013teleparallel}
\begin{equation}\label{eq33}
\phi_{ab}=\frac{1}{2}\Big( T_{(0)ab} + T_{a(0)b} - T_{b(0)a} \Big)\,.
\end{equation}
The components $\phi_{(0)(i)}$ give the inertial accelerations $\mathbf{a}$ of the frame, while $\phi_{(i)(j)}$ represents the rotation $\mathbf{\Omega}$ of the frame relative to a non-rotating Fermi-Walker transported frame \cite{mashhoon2002length,mashhoon2003vacuum}. This means that different frames (observers) can be present in a spacetime, and consequently, distinct tetrads can be constructed for the same metric tensor.

If a frame is static, then $\lim_{r\rightarrow\infty}e^{a}\,_{\mu}=\delta^{a}_{\mu}$ holds for an asymptotically flat spacetime, allowing us to define a static frame with respect to a static frame at infinity.

The evaluation of the irreducible energy requires an observer located very close to the event horizon, inside the ergosphere. Inside this region, no static observer can exist. Therefore, we need a tetrad field associated with the metric (\ref{eq21}) and possessing longitudinal rotation, meaning that its four-velocity is given by $u^{\mu}=(u^{0},0,0,u^{3})$ with $u^{3}\neq0$. One set of tetrads satisfying these conditions is
\begin{equation}\label{eq34}
e_{a\mu}=K\,\left(
\begin{array}{cccc}-A&0&0&0\\
B\sin\theta\sin\phi
&C\sin\theta\cos\phi& D\cos\theta\cos\phi&-G\sin\theta\sin\phi\\
-B\sin\theta\cos\phi
&C\sin\theta\sin\phi& D\cos\theta\sin\phi& G\sin\theta\cos\phi\\
0&C\cos\theta&-D\sin\theta&0
\end{array}
\right)\,,
\end{equation}
where
\begin{align}
A&= {1\over \rho}\sqrt{\psi^2+{\chi^2\over \Sigma^2}\sin^2\theta}\,,
\nonumber \\
B&={\chi \over {\Sigma \rho}}\,, \nonumber \\
C&={\rho \over \sqrt{\Delta}}\,, \nonumber \\
D&=\rho\,, \nonumber \\
G&= {\Sigma \over \rho}\,. \nonumber \\
K&=\sqrt{\frac{\rho^{2}+b/r^{3}}{\rho^{2}}}\,.\label{eq35}
\end{align}
The tetrads (\ref{eq34}) yields the determinant
$
e=K^4\sqrt{\frac{ \chi^2 \sin ^2\theta  +\psi^2\Sigma^{2}}{\Delta}}\,\sin \theta\,,
$
and the four velocity of the observer is
\begin{equation}\label{eq37}
u^{\mu}=e_{(0)}\,^{\mu}=\frac{\rho\,\Sigma}{K\,\sqrt{\Sigma^{2}\phi^{2}+\chi^{2}\sin^{2}\theta}}\Big(1 ,0,0,\chi/\Sigma \Big)\,.
\end{equation}
From (\ref{eq37}), we can see that the observer rotates around the black hole with angular velocity $\omega=u^{3}$ with respect to a static observer at infinity. The angular velocity is zero at infinity.

Given a metric tensor and an associated tetrad, when the parameters of the metric tensor (mass, angular momentum, etc.) go to zero, the resulting tetrad, in Cartesian coordinates, which neither rotates nor translates with respect to Minkowski space is the one in which $e_{(i)j} = e_{(j)i}$ and $e_{(0)}\,^{i} = 0$, and for practical purposes discussed in the article, we are using this type of tetrad. This is important to avoid inertial/dynamic effects carried by measurements made by observers adapted to these tetrads.

Using the tetrads (\ref{eq34}) with the definitions (\ref{eq35}), the non-zero components of the torsion tensor can be evaluated as

\begin{minipage}{.5\textwidth}
\begin{align}
T_{(0)01} &= \partial_{r}\left( \frac{\sqrt{\Psi}}{\rho^2} \sqrt{\psi^2 +\chi^2 \sin^2 (\theta)/\Sigma^2} \right)\,,\nonumber\\
T_{(0)02} &= \partial_\theta \left( \frac{\sqrt{\Psi}}{\rho^2} \sqrt{\psi^2 +\chi^2 \sin^2 (\theta)} \right)\,,\nonumber\\
T_{(1)01} &= - \partial_r \left( \frac{\sqrt{\Psi}}{\rho^2} \frac{\chi}{\Sigma}\right) \sin(\theta)\sin(\phi) \,,\nonumber\\
T_{(1)02} &= - \partial_\theta \left( \frac{\sqrt{\Psi}}{\rho^2} \frac{\chi}{\Sigma} \sin(\theta)\right)\sin(\phi)\,,\nonumber\\
T_{(1)03} &= \frac{\sqrt{\Psi}}{\rho^2} \frac{\chi}{\Sigma}\sin(\theta)\cos(\phi)\,,\nonumber\\
T_{(1)12} &= \left(\partial_r\sqrt{\Psi} \cos(\theta) - 
\partial_\theta\left(\frac{\sqrt{\Psi}}{\Sigma} \sin(\theta)\right)\right)\cos\phi\,,\nonumber\\    
T_{(1)13} &= \left(\frac{\sqrt{\Psi}}{\rho^2}\Sigma - 
\frac{\sqrt{\Psi}}{\Delta}\right)\sin(\theta)\sin(\phi)\,,\nonumber\\
T_{(1)23} &= \left(-\partial_\theta \left(\frac{\sqrt{\Psi}}{\rho^2}\Sigma\sin(\theta)\right) + 
\sqrt{\Psi}\cos(\theta)\right)\sin(\phi)\,,\nonumber
\end{align}
\end{minipage}
\begin{minipage}{.5\textwidth}
\begin{align}
T_{(2)01} &= \partial_r\left(\frac{\sqrt{\Psi}\chi}{\rho^2\Sigma}
\right)\sin(\theta)\cos(\phi)\,,\nonumber\\
T_{(2)02} &= \partial_\theta \left(\frac{\sqrt{\Psi}\chi}{\rho^2\Sigma}
\sin(\theta)\right)\cos(\phi)\,,\nonumber\\
T_{(2)03} &= -\frac{\sqrt{\Psi}\chi}{\rho^2\Sigma}
\sin(\theta)\cos(\phi)\,,\nonumber\\
T_{(2)12} &= \left(\partial_r\sqrt{\Psi} \cos(\theta) - 
\partial_\theta\left(\frac{\sqrt{\Psi}}{\sqrt\Delta}\sin(\theta)\right)\right)\sin(\phi)\,,\nonumber\\ 
T_{(2)13} &= \left(\partial_r\left(\frac{\sqrt{\Psi}}{\rho^2\Sigma}\right) -
\frac{\sqrt{\psi}}{\sqrt\Delta}\right)\sin(\theta)\cos\phi\,,\nonumber\\
T_{(2)23} &= \left(\partial_\theta\left(\frac{\sqrt{\Psi}}{\rho^2\Sigma}\sin(\theta)\right) -
\sqrt{\psi}\cos(\theta)\right)\cos\phi\,,\nonumber\\
T_{(3)12} &= -\left(\partial_r\sqrt{\Psi}\sin(\theta) + 
\partial_\theta\left(\frac{\sqrt{\Psi}}{\sqrt{\Delta}}\cos(\theta)\right)\right)\,,\label{eq38}
\end{align}
\end{minipage}
%where the prime indicates a derivative with respect to $r$.
\vspace{.1cm}
where $\Psi=\rho^{2}K^{2}$.

From the torsion tensor components (\ref{eq38}), we can calculate the relevant projected component $\Sigma^{(0)01}$ of the superpotential (\ref{eq7}). The calculations, although lengthy, are straightforward, and we obtain
\begin{equation}\label{eq39}
\Sigma^{(0)01}=
\Sigma \sqrt{\Delta} \frac{ K \left(\rho^2+\Sigma\right)-\sqrt{\Delta}\Big(K \partial_{r} \Sigma+2  \Sigma \partial_{r} K \Big) }{2 \rho K^4 \left(\Sigma^2 \psi^2+\sin ^2\theta \chi^2\right)}\sqrt{\frac{\sin ^2\theta \chi^2}{\Sigma^2}+\psi^2}\,.
\end{equation}

The total energy of the gravitational field can be evaluated either from the volume integral in equation (\ref{eq15}) or from the surface integral. The surface integral approach is more advantageous, as we can choose a surface $dS_1$ with a constant radius $r=R$ that encompasses the entire black hole. Thus, for the total energy $E$ inside a surface with $R$ constant, we have
\begin{align}
E = P^{(0)} &= 4k\oint_{S} dS_1 \left(e\,\Sigma^{(0)01}\right) \nonumber \\
&= 4k \lim_{r\rightarrow R} \int_{0}^{2\pi} d\phi \int_{0}^{\pi} d\theta \left(e\,\Sigma^{(0)01}\right), \label{eq40}
\end{align}
where $\Sigma^{(0)01}$ is given by equation (\ref{eq39}). For a surface at infinity, we have $E=M$ regardless the values of the parameters. In other words, as $r$ approaches infinity, an observer measures the same energy as that of the Schwarzschild black hole, similar to what happens with the Kerr black hole.
Numerical integration can be performed for any surface $R>r_+$, allowing us to compare the effects of the additional parameters $e$ and $b$ with the Kerr solution. By integrating the energy (\ref{eq40}) over several surfaces and tracking the resulting interpolation curve, we can analyze the behavior and compare it with the Kerr solution.
\begin{figure}
\centering
\begin{minipage}{.47\textwidth}
	\centering
		\includegraphics[width=1\textwidth]{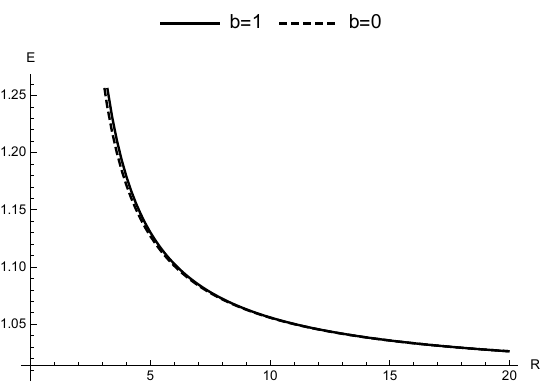}
	\caption{Numerical integration of the energy (\ref{eq40}) for surfaces of distinct radii $R>r_{+}$. The outer horizon is at $r_{+}=1.99499$, and the parameters are $M=10$, $a=1$, and $e=0$. The continuous line represents the regular black hole ($b=1$), and the dotted one represents the Kerr black hole ($b=0$).}
	\label{fig1}
\end{minipage}
\qquad
\begin{minipage}{.47\textwidth}
	\centering
		\includegraphics[width=1\textwidth]{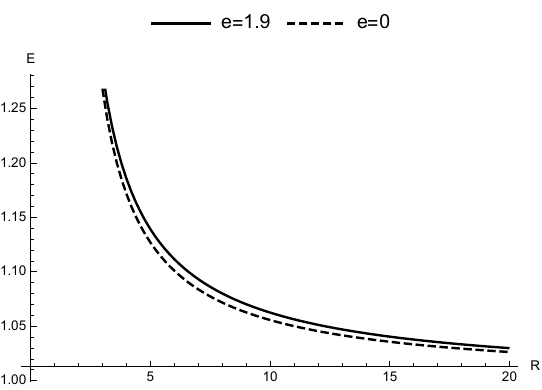}
	\caption{Numerical integration of the energy (\ref{eq40}) for surfaces of distinct radii $R>r_{+}$. The outer horizon is at $r_{+}=1.99499$, and the parameters are $M=10$, $a=1$, and $b=0$. The continuous line represents the regular black hole ($e=1.9$), and the dotted one represents the Kerr black hole ($e=0$).}
	\label{fig2}
\end{minipage}
\end{figure}

In Figure \ref{fig1}, we present the comparison between a regular black hole and the Kerr black hole, considering specific parameter values. The regular black hole is characterized by a deviation parameter $e$ set to zero and a regularization parameter $b$ set to unity. The continuous curve represents the regular black hole, while the dashed curve represents the Kerr black hole. It is noteworthy that in the vicinity of the outer horizon, the energy of the regular black hole is slightly higher than that of the Kerr black hole. However, as the surface radius increases, the energies converge and become indistinguishable.
Furthermore, we investigate the effect of varying the deviation parameter $e$ while keeping the regularization parameter $b$ fixed at zero, within the range specified by equation (\ref{eq32}). For most values within this range, the differences in energy are negligible. However, as we approach the inner horizon of the outer one, i.e., when $e$ tends towards 2, a significant change in energy is observed. To illustrate this extreme condition, we provide a plot of the energies in Figure \ref{fig2}.

%%%%%%%%%%%%%%%%%%%%%%%%%%%%%%%%%%%%%%%%%%%%%%%%%%%%%%%%%%%%%%%%%%%%%%%%%%%%%%%%%%%%%%%%%%%%%%%
\subsection{The inertial accelerations close do the inner horizon}
%%%%%%%%%%%%%%%%%%%%%%%%%%%%%%%%%%%%%%%%%%%%%%%%%%%%%%%%%%%%%%%%%%%%%%%%%%%%%%%%%%%%%%%%%%%%%%%

As discussed in Section \ref{review}, the presence of only the regularization parameter $b$ leads to a non-zero surface gravity near the inner horizon, resulting in the mass inflation problem characterized by a specific geodesic behavior of particles in close proximity to this horizon.

The tetrads carry essential information about both the spacetime and the observer's frame, allowing us to analyze the dynamic behavior of the vicinity of the inner horizon by evaluating the inertial accelerations experienced by an observer in that region. The inertial accelerations given by equation (\ref{eq33}) ensure that the observer remains in the kinematic state compatible with 4-velocity (\ref{eq37}) in the presence of the gravitational field. Therefore, by assessing the inertial accelerations of an observer, we can gain insights into the acceleration properties of the gravitational field itself.

Although Boyer-Lindquist coordinates are regular everywhere outside the outer horizon, they are not suitable for investigating the region between the inner and outer horizons of a black hole. The tetrads defined in equation (\ref{eq34}), on the other hand, possess both spacetime and local indices, behaving as four-vectors under coordinate transformations. While the physical quantities projected onto the local frame of an observer remain regular throughout the spacetime, the limitations of the coordinate system inside the event horizon of a black hole make the tetrads defined in equation (\ref{eq34}) unsuitable for investigating the region between the horizons, as they become complex.

However, the tetrads given by equation (\ref{eq34}) remain regular inside the inner horizon, allowing us to evaluate the inertial accelerations of an observer (\ref{eq34}) as they "approach" the inner horizon from the inside.
From equation (\ref{eq33}), we can observe that the inertial accelerations can be obtained by projecting the torsion tensor as follows:
\begin{equation}\label{eq41}
a_{(i)} \equiv \phi_{(0)(i)} = T_{(0)(0)(i)}.
\end{equation}
In order to simplify our analysis, we consider an observer located in the equatorial plane. By projecting the torsion tensor components (\ref{eq38}) into the local frame, specifically for $\theta=\pi/2$, we obtain
\begin{equation}\label{eq42}
\vec{a}=a\, \cos{\phi} \hat{x} + a\, \sin{\phi} \hat{y}\,,
\end{equation}
where
\begin{equation}\label{eq43}
a=\frac{\sqrt{\Delta}}{K^2 \rho^2} \left[\rho\partial_r K  + K \left(\frac{\rho \left(-\chi^2 \partial_r \Sigma + \Sigma \chi \chi'  + \Sigma^3 \psi \partial_r \psi\right)}{\Sigma \left(\Sigma^2 \psi^2 + \chi^2\right)} - \partial_r \rho\right)\right]\,.
\end{equation}
By defining $\hat{r}=\cos{\phi}\hat{x}+\sin{\phi}\hat{y}$ in the equatorial plane, we obtain the inertial acceleration
\begin{equation}\label{eq44}
\vec{a}=a\,\hat{r}\,.
\end{equation}
The quantity (\ref{eq43}) is consistently positive outside the outer horizon, as anticipated. A positive inertial acceleration implies an inward gravitational acceleration. Inside the inner horizon, the inertial acceleration (\ref{eq43}) becomes negative, indicating a gravitational repulsion originating from the interior of the black hole.

While the Kerr deviation parameter has minimal effect on the energy of the black hole, as depicted in Figure \ref{fig2}, it significantly influences the qualitative behavior of the inertial acceleration.
\begin{figure}
\centering
\begin{minipage}{.47\textwidth}
	\centering
		\includegraphics[width=1\textwidth]{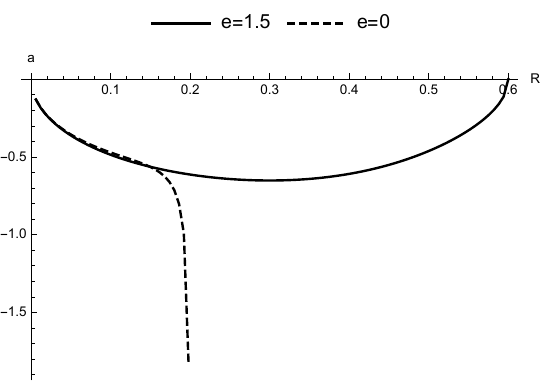}
	\caption{Inertial acceleration of the observer inside the inner horizon, comparing the cases with a Kerr deviation parameter (continuous line) and without a Kerr regularization parameter (dashed line). The chosen parameters for the plot are $M=10$, $a=0.6$, and $b=0.5$.}
	\label{fig3}
\end{minipage}
\qquad
\begin{minipage}{.47\textwidth}
	\centering
		\includegraphics[width=1\textwidth]{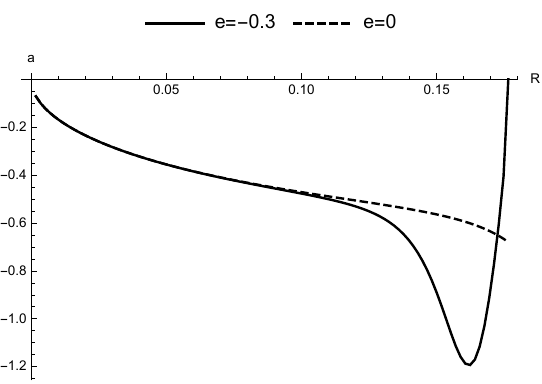}
	\caption{Inertial acceleration of the observer inside the inner horizon, with the Kerr deviation parameter represented by the continuous line and without the Kerr regularization parameter shown by the dashed line. The chosen parameters for the plot are $M=10$, $a=0.6$, and $b=0.5$.}
	\label{fig4}
\end{minipage}
\end{figure}
In Figure \ref{fig3}, we present the plot of the inertial acceleration (\ref{eq43}) within the black hole for various values of $r=R$. It is evident that the inertial acceleration exhibits an exponential increase in magnitude when the regularization parameter $e$ is not taken into account, as illustrated by the dashed lines. However, when the deviation parameter is considered (continuous line), the inertial acceleration tends towards zero as the observer approaches the inner horizon, indicating that the gravitational acceleration in close proximity to that region vanishes. Similarly, in Figure \ref{fig4}, we observe the same behavior for a negative Kerr deviation parameter, corresponding to a contracted inner horizon.

%%%%%%%%%%%%%%%%%%%%%%%%%%%%%%%%%%%%%%%%%%%%%%%%%%%%%%%%%%%%%%%%%%%%%%%%%%%%%%%%%%%%%%%%%%%%%%%%%%%%%%%%%%%%%%%%%%%%%%%%%%%%%%%
%%%%%%%%%%%%%%%%%%%%%%%%%%%%%%%%%%%%%%%%%%%%%%%%%%%%%%%%%%%%%%%%%%%%%%%%%%%%%%%%%%%%%%%%%%%%%%%%%%%%%%%%%%%%%%%%%%%%%%%%%%%%%%%
\subsection{The irreducible mass}\label{mass}

The interaction between a black hole and a particle can be studied using perturbation methods. However, when considering a particle with a much smaller mass compared to the black hole's parameter $M$, we can neglect the particle's geometrical influence on spacetime. In the case of a Kerr black hole, if such a particle decays inside the ergosphere, with one part falling into the outer horizon and the other escaping from the ergosphere, the escaping particle can have greater energy than the initial particle, as demonstrated by Penrose. In the absence of any other entities besides the black hole and the particle, it becomes apparent that the black hole has lost rotational energy to the particle.

In 1970, Christodoulou \cite{christodoulou1970reversible} showed that there exists a maximum energy that a particle can extract from a Kerr black hole, leaving behind a remaining mass known as the irreducible mass of the black hole. The physical mass of a black hole is measured at spatial infinity, resulting in the expression $M$ in Equation (\ref{eq40}). In the Schwarzschild limit with $a=0$, the energy at the event horizon is given by $E=2M$.

The outer horizon of a Kerr black hole is an event horizon, meaning that the energy contained within it cannot escape. Therefore, regardless of the number of particles with which the black hole interacts, its energy cannot decrease below the energy contained within the outer horizon prior to the interactions. We can compute the energy contained within the outer horizon as the irreducible energy $E_{irr}$ of the black hole and compare it with the relation
\begin{equation}\label{eq45}
E_{irr}=2\,M_{irr}\,,
\end{equation}
where we identify $M_{irr}$ as the irreducible mass of the Kerr black hole. This procedure for the Kerr black hole yields numerical values that are nearly identical to those obtained by Christodoulou's method \cite{maluf1996gravitational}.

The regular rotating black hole, as described by Equation (\ref{eq21}), also possesses an ergosphere. Consequently, we proceed with the evaluation of its irreducible mass. To accomplish this, we need to calculate the surface integral given by Equation (\ref{eq40}) at the outer horizon. Taking the limit as we approach the outer horizon, i.e.,
\begin{equation}\label{eq46}
\displaystyle{\lim_{r\rightarrow r_{+}}\Big(e,\Sigma^{(0)01}\Big)}=K\, \frac{\rho^{2}+\Sigma}{2\rho}\sin{\theta}\\,
\end{equation}
we obtain the expression for the energy as
\begin{equation}\label{eq47}
P^{(0)}=\left.\frac{1}{4}\int_{0}^{\pi}\sin{\theta}d\theta \, \left[\frac{(2r^{2}+a^{2}+a^{2}\cos^{2}\theta)\sqrt{1+\frac{b}{r^{5}+r^{3}a^{2}\cos^{2}{\theta}}}}{\sqrt{r^{2}+a^{2}\cos^{2}{\theta}}}\right]\,\right|_{r=r{+}}\,.
\end{equation}
Fortunately, this integral (\ref{eq47}) can be analytically evaluated. By introducing the quantities $\Sigma_{+}=(r_{+}^{2}+a^{2})$, $\alpha=\sqrt{r_{+}^{2}+b/r_{+}^{3}}$, and $\beta=\alpha/r_{+}$, we can express the irreducible energy of the rotating regular black hole as
\begin{equation}\label{eq48}
E_{irr}=\frac{\alpha}{4a}\Bigg[ a \sqrt{1+\frac{a^{2}}{\alpha^{2}}} - \alpha \left( \frac{1}{2}+\frac{\Sigma_{+}}{\alpha^{2}} \right) \ln\left( \frac{\sqrt{a^{2}+\alpha^{2}}-a}{\sqrt{a^{2}+\alpha^{2}}+a} \right) +2 \frac{\Sigma_{+}\sqrt{\beta^{2}-1}}{\alpha}\tan^{-1}\left( \frac{\sqrt{\beta^{2}-1}}{\sqrt{a^{2}+\alpha^{2}}}a \right) \Bigg]\,.
\end{equation}
Hence, from the identity (\ref{eq45}) and from (\ref{eq48}), we obtain the main result of this article, i.e., the irreducible mass of the regular rotating black hole,
\begin{equation}\label{eq49}
M_{irr}=\frac{2 {a} \sqrt{r_{+}^3 \left(b+{a}^2 r_{+}^3+r_{+}^5\right)}-\left(b+2 {a}^2 r_{+}^3+3 r_{+}^5\right) \ln
   \left(\frac{\sqrt{{a}^2+\frac{b}{r_{+}^3}+r_{+}^2}-{a}}{{a}+\sqrt{\mathit{
   a}^2+\frac{b}{r_{+}^3}+r_{+}^2}}\right)+4 \left(\sqrt{b r_{+}^5}+{a}^2 \sqrt{b r_{+}}\right)
   \tan ^{-1}\left(\frac{{a} \sqrt{\frac{b}{b+{a}^2 r_{+}^3+r_{+}^5}}}{r_{+}}\right)}{16
   {a} r_{+}^3}\,.
\end{equation}
An important observation regarding Equation (\ref{eq49}) is the absence of the Kerr deviation parameter $e$. Consequently, the irreducible mass is solely dependent on the mass, rotation, and regularization parameters. In equations (\ref{eq26}-\ref{eq29}), we carefully selected the parameters $\alpha, \beta, \gamma, \mu$ to ensure that the outer horizon (\ref{eq30}) coincides with the Kerr horizon. As a result, it is expected that the energy contained within the outer horizon, and hence the irreducible mass, does not rely on the position of the inner horizon determined by $e$. It is worth noting that the method employed here to calculate the irreducible mass of the black hole is not unanimous. There may even be objections to the generalization of its use, since it has not been demonstrated that such a method has general validity. %On the other hand, the method is equivalent to that used by Christodoulou for the Kerr black hole, in addition to generating the correct limits for the regular black hole.
\begin{figure}
\centering
\begin{minipage}{.47\textwidth}
	\centering
		\includegraphics[width=.9\textwidth]{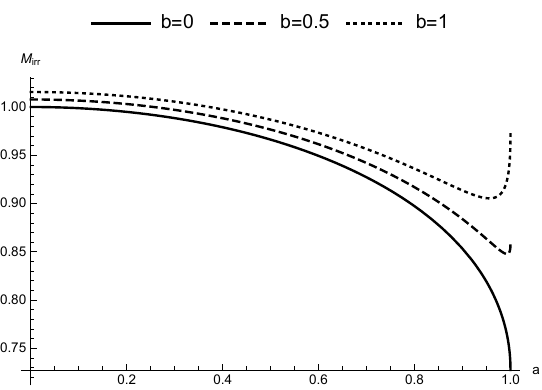}
	\caption{Irreducible mass as a function of $a$ for different values of the regularization parameter $b$.}
	\label{fig5}
\end{minipage}
\qquad
\begin{minipage}{.47\textwidth}
	\centering
		\includegraphics[width=.9\textwidth]{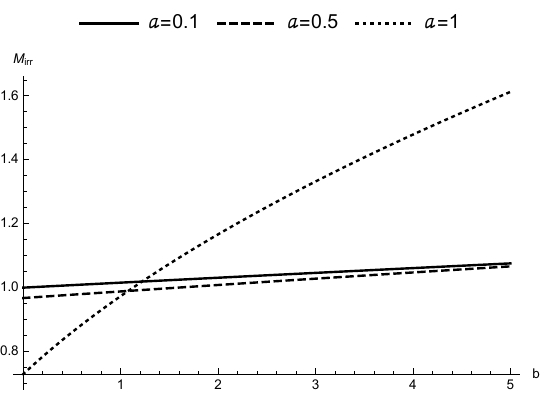}
	\caption{Irreducible mass as a function of $b$ for different values of the rotation parameter $a$.}
	\label{fig6}
\end{minipage}
\end{figure}

Figure \ref{fig5} illustrates the behavior of the irreducible mass (\ref{eq49}) as a function of $a$, considering three distinct values of the regularization parameter $b$. The continuous line represents the Kerr irreducible mass, which yields the same result as in Ref. \cite{maluf1996gravitational}. When a non-zero value is assigned to $b$, a similar pattern is observed for lower values of $a$, but discrepancies become more pronounced as $a$ approaches unity. Prior to reaching unity, the irreducible mass begins to increase. Consequently, there exists a minimum value for the irreducible mass that shifts to the left as the value of $b$ increases. This can be observed by comparing the dotted and dashed lines in Figure \ref{fig5}.

Figure \ref{fig6} depicts the behavior of (\ref{eq49}) for fixed values of $a$. For most values of $a$, the behavior is comparable. However, as $a$ approaches unity, the behavior undergoes significant changes, gradually approaching the dotted line in Figure \ref{fig6}, which represents the extreme case.

%%%%%%%%%%%%%%%%%%%%%%%%%%%%%%%%%%%%%%%%%%%%%%%%%%%%%%%%%%%%%%%%%%%%%%%%%%%%%%%%%%%%%%%%%%%%%%%%%%%%%%%%%%%%%%%%%%%%%%%%%%%%%%%
%%%%%%%%%%%%%%%%%%%%%%%%%%%%%%%%%%%%%%%%%%%%%%%%%%%%%%%%%%%%%%%%%%%%%%%%%%%%%%%%%%%%%%%%%%%%%%%%%%%%%%%%%%%%%%%%%%%%%%%%%%%%%%%
\section{Conclusions}\label{conclusions}

Black hole mimickers are significant astrophysical objects that exhibit observational properties nearly identical to those of regular black holes \cite{lemos2008black} while lacking physical singularities. Observations in the Sagittarius A region of the Milky Way have revealed shadow-like features that resemble those predicted to occur around a Kerr black hole, suggesting the presence of a supermassive black hole at the center of our galaxy. Recent data reanalysis indicates that the same observed shadow of Sagittarius A can be expected in the presence of a regular rotating black hole \cite{shaikh2023testing}. This finding emphasizes the importance of considering these mathematical objects as plausible candidates for explaining the experimental data associated with black holes, thereby elevating them to the status of real astronomical entities.

In general, all regular black holes necessitate a source with energy-momentum that violates one or more energy conditions, similar to the Alcubierre solution and wormholes. However, energy conditions are not physical laws but rather expected behavior in nature. Therefore, the existence of such objects is no more unreasonable than the presence of actual physical singularities.

Given the significance of this topic and the proximity of potential candidates, our study focuses on the irreducible mass of a regular rotating black hole. If these objects indeed exist in nature, their interaction with particles and consequent energy extraction through the Penrose process are inevitable. By adopting the procedure outlined in Ref. \cite{maluf1996gravitational} and employing the energy-momentum definitions of TEGR, we evaluated the energy density of the regular black hole (\ref{eq21}) and numerically computed its total energy for different surfaces outside the outer horizon. 
The obtained pattern is very much similar to that of a Kerr black hole. Thus, we can conclude that the regular black hole (\ref{eq21}) closely resembles the Kerr black hole, except in a very close range to the outer horizon. Additionally, we demonstrated that the total energy is minimally influenced by the Kerr deviation parameter, indicating that the displacement of the inner horizon relative to the Kerr horizon has little effect on the overall energy of the black hole.

We also examined the acceleration tensor within the inner horizon and compared the cases of zero deviation from Kerr inner horizon to a non-zero deviation. Our analysis revealed an infinite gravitational acceleration as the observer approaches the inner horizon, while considering the deviation parameter yielded a value of zero. This finding reinforces the significance of the parameter $e$ in stabilizing the inner horizon and resolving the mass-inflation problem.

Following the interior analysis, we proceeded with the primary focus of our article, which was to determine the irreducible mass of the regular black hole (\ref{eq21}). By taking the limit of the outer horizon in the surface integral (\ref{eq40}), we were able to analytically integrate the energy at that boundary and obtain the energy contained within the event horizon (\ref{eq48}). We confirmed that this expression reduces to (\ref{irr_maluf}) of the Kerr black hole in the limit $b\rightarrow 0$. Moreover, by utilizing the relation (\ref{eq45}), we identified the irreducible mass (\ref{eq49}) of the regular black hole (\ref{eq21}). Although the gravitational energy and the irreducible mass obtained in the article, equations (\ref{eq48}) and (\ref{eq49}), may depend on inertial effects of observers, for the analysis conducted in the article, we chose observers adapted to tetrads that do not carry such effects in the Minkowski limit. Additionally, the irreducible mass for the Kerr black hole obtained by Chistodoulou is not obtained through a volume/surface integral but rather through calculations along the equatorial plane of the Kerr black hole, which may explain the small difference compared to the result obtained by us.

From equation (\ref{eq49}), it is evident that the presence of the regularization parameter $b$ induces a qualitative change in the behavior of the irreducible mass, as depicted in Figure \ref{fig5}. It is noteworthy that for each regular black hole, characterized by a specific value of $b$, there exists a corresponding value of $a$ that allows for the maximum extraction of energy from the black hole, resulting in a minimum irreducible mass. This critical value of $a$ is extremely close to the extreme case $a=1$, but it can be attained for lower values of $a$ if the regularization parameter is sufficiently large.

Since the energy (\ref{eq48}) encompasses the total energy (source plus gravitational field) of the black hole, it is possible to explore the thermodynamics of such objects as unified thermodynamic systems. However, a detailed analysis of this aspect will be pursued in future investigations.

%%%%%%%%%%%%%%%%%%%%%%%%%%%%%%%%%%%%%%%%%%%%%%%%%%%%%%%%%%%%%%%%%%%%%%%%%%%%%%%%%%%%%%%%%%%%%%%%%%%%%%%%%%%%%%%%%%%%%%%%%%%%%%%
%%%%%%%%%%%%%%%%%%%%%%%%%%%%%%%%%%%%%%%%%%%%%%%%%%%%%%%%%%%%%%%%%%%%%%%%%%%%%%%%%%%%%%%%%%%%%%%%%%%%%%%%%%%%%%%%%%%%%%%%%%%%%%%

\end{document}